# Breaking the trade-off between invisibility and sensitivity in electromagnetic sensing

Yichao Liu [1], Jiaxue Zhou [1], Weifeng Han [1], Hanchuan Chen [1], Fei Sun [1]*, Qin Liao [1], Hengxiang Zhang [1], Xiaofan Ji [1], Yawen Qi [1]

[1] *Key Lab of Advanced Transducers and Intelligent Control System, Ministry of Education and Shanxi Province, College of Physics and Optoelectronics, Taiyuan University of Technology, Taiyuan, 030024 China;*

sunfei@tyut.edu.cn

**Abstract**: Weak electromagnetic signals demand highly sensitive sensors, yet increasing a sensor's sensitivity inevitably strengthens its interaction with the surrounding field, producing scattering that perturbs the very signals being measured. Conversely, existing cloaking strategies suppress scattering only by isolating the sensor from incident waves, thereby compromising signal reception. Resolving this long-standing trade-off between invisibility and sensitivity has remained an outstanding challenge. Here we overcome this dilemma through an integrated transformation-optical architecture that co-designs the entire sensing system, including the electrically large sensor body, the subwavelength sensing probe, and their electrical interconnection. The proposed multifunctional core–shell structure guides incident waves around the sensor body while simultaneously concentrating them into the sensing region without disturbing the external electromagnetic field. A deep-subwavelength aperture preserves electrical connectivity without degrading either cloaking or field concentration, enabling invisible sensing within a single platform. A microwave prototype based on practical optic-null-medium metamaterials experimentally demonstrates broadband scattering suppression exceeding 3 dB together with an average sixfold enhancement of the detected signal over 4.9–5.1 GHz. By simultaneously eliminating measurement-induced field perturbation and amplifying the local sensing field, our approach establishes a general framework for invisible yet highly responsive electromagnetic sensors, opening new opportunities for weak-signal detection in biomedical diagnostics, secure communications, quantum technologies, and deep-space exploration.

## 1 Introduction

The detection of extremely weak electromagnetic (EM) signals—such as those emanating from neural circuits at powers as low as $10^{-16}$ W—is essential for a broad range of emerging technologies, including neuroscience [1], quantum technology [2, 3], secure communications, biomedical diagnostics, and deep-space exploration [4, 5]. Achieving reliable detection of such weak signals requires sensors with extremely high sensitivity. However, increasing a sensor's sensitivity inevitably strengthens its interaction with the surrounding EM field, introducing scattering and absorption that perturb the very signals it is intended to measure. This measurement back-action creates a fundamental sensing paradox: amplifying the desired signal inevitably amplifies sensor-induced field distortion, thereby limiting detection fidelity. Consequently, high sensitivity and electromagnetic invisibility become seemingly incompatible objectives—a long-standing challenge that has hindered the development of non-invasive weak-signal detection technologies.

Considerable efforts have been devoted to addressing the two requirements of sensitivity enhancement and electromagnetic invisibility independently. On the one hand, sensor sensitivity has been improved through resonant metamaterials, field-concentrating structures, and optimized sensing elements, such as split-ring resonators and folded structures, which locally enhance the EM field and strengthen signal reception [6–8]. On the other hand, sensor invisibility has been pursued using a variety of cloaking techniques, including scattering-cancellation cloaks based on plasmonic shells or metasurface covers [9–14], as well as ray-bending cloaks enabled by transformation optics (TO) [16–23]. Despite their remarkable progress, these two research directions have evolved largely in parallel. Sensitivity-enhancement strategies intentionally reinforce EM interactions, whereas cloaking techniques suppress or redirect them, leaving the fundamental trade-off between sensitivity and invisibility unresolved.

A key practical reason for this long-standing challenge is that existing studies generally treat a sensor as an isolated sensing probe, whereas a functional sensor is in fact an integrated system consisting of a subwavelength sensing probe, an electrically large main body that accommodates signal-processing and power modules, and the electrical connection between them. These components introduce two fundamentally different perturbation mechanisms: scattering directly produced by the electrically large main body and absorption-induced scattering associated with the subwavelength probe. Existing cloaking strategies can effectively suppress only one of these mechanisms [15]. Ray-bending cloaks based on TO [16–23] conceal the main body but isolate the sensor from incident waves, leading to the "mutually blind" condition [9] that prevents signal detection. Conversely, scattering-cancellation cloaks [24–28] reduce the probe's scattering while remaining ineffective against the dominant scattering generated by the electrically large sensor body. Consequently, no existing approach can simultaneously suppress both scattering mechanisms while preserving efficient signal reception. An ideal invisible sensor should therefore render the entire sensing system—including both the main body and the sensing probe—electromagnetically invisible, while simultaneously enhancing the local EM field at the probe and maintaining their electrical interconnection. Achieving this system-level co-design remains a fundamental challenge for non-invasive weak-signal detection.

Here, we resolve this long-standing trade-off between electromagnetic invisibility and sensing sensitivity through an integrated TO strategy that co-designs the entire sensing system rather than its individual components [29–31]. Specifically, we propose a multifunctional core–shell (MCS) architecture based on optic-null media (ONM) [32–36]. Without the MCS (Fig. 1A, see also **Movie 1**), scattering from both the electrically large sensor body and the subwavelength sensing probe perturbs the surrounding EM field, attenuates the received signal, and compromises measurement accuracy. In contrast, with the MCS (Fig. 1B), these conflicting requirements are simultaneously resolved through a unified physical framework. The proposed MCS consists of an ONM shell enclosing multiple concealed regions for the sensor's main body and a central detecting region formed by a high-refractive-index (HRI) core that accommodates the sensing probe. Deep-subwavelength apertures are strategically introduced between the concealed regions and the detecting region to preserve continuous electrical connectivity between the sensor body and the probe without compromising either functionality. The ONM shell guides incident EM waves around the concealed regions while simultaneously tunneling them into the HRI core without reflection or phase delay, thereby suppressing scattering from the sensor body. Meanwhile, a continuous squeezing transformation concentrates the incident EM field inside the HRI core, selectively enhancing the local field at the sensing probe while preserving the external wavefront. This dual mechanism simultaneously realizes system-level cloaking and field enhancement, thereby reconciling invisibility and sensitivity within a single transformation-optical platform. Moreover, the concealed regions can accommodate varied body geometries, numbers, sizes, and materials, offering considerable design flexibility.

Using practical ONM metamaterials, we experimentally demonstrate this invisible sensing architecture at microwave frequencies. The fabricated prototype exhibits broadband scattering suppression together with an approximately sixfold enhancement of the detected signal. Beyond this specific implementation, our work establishes a general framework for system-level co-design of invisible yet highly responsive EM sensors, providing a scalable strategy for non-invasive weak-signal detection in biomedical diagnostics, secure communications, quantum technologies, and deep-space exploration.

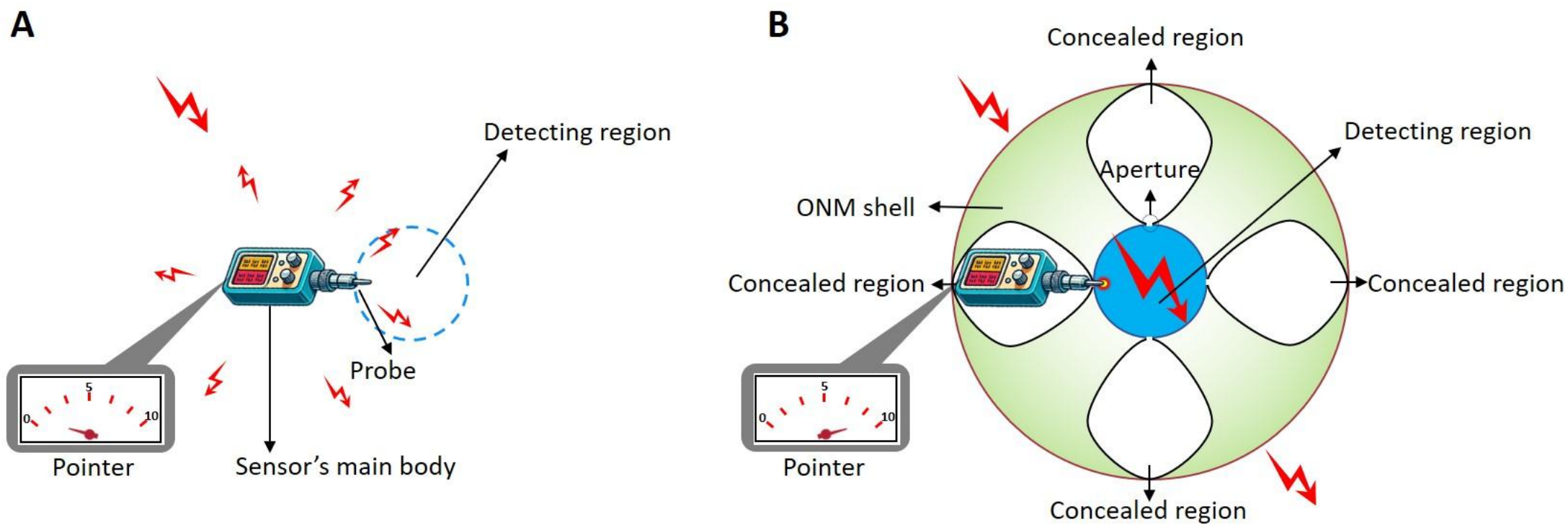

**Fig. 1. Sensor operation for weak EM field detection (A) without and (B) with the MCS.** (**A**) A bare sensor shows negligible response, and its presence severely disturbs the ambient EM field due to scattering. (**B**) With the MCS, the external field remains unperturbed, while the field around the probe is locally amplified, producing a clear sensor response.

# 2 Methods

## 2.1 Design of the two-dimensional MCS Structure

To illustrate the design approach, two-dimensional (2D) design in the cylindrical coordinate system is first utilized to demonstrate the three-step design process of the MCS. The first step involves the design of a ONM-based EM concentrator by using TO, as depicted in Fig. 2A to Fig. 2B. This process includes stretching the inner boundary of the purple region (i.e., a circle with a radius of $R_3$-D, as shown in Fig. 2A) to a circle with a radius of $R_1$, while the outer boundary of the purple region (i.e., a circle with a radius of $R_3$) remains fixed **(Supplemental Note 1)**. Correspondingly, the white region is compressed into a circular area with a radius of $R_1$. After the transformation with $\Delta\to 0$, a concentrator is created as shown in Fig. 2B, which consists of two parts, i.e., the HRI core (colored blue) and the ONM shell (colored green). The HRI core has a refractive index of $n_c = R_3/R_1$, and the ONM shell has its permittivity and permeability approaching infinity along its principal axes (i.e., the radial direction, indicated by the blue arrows in Fig. 2B) and equal to zero in the other two perpendicular directions. This allows incident waves to be guided without reflection or phase delay into the HRI core [36], enhancing the field while leaving the background undisturbed.

The second step involves creating air regions to conceal the sensor's main body. A secondary transformation within the ONM squeezes the material along a circle of radius $R_2$, creating four eye-shaped air regions (colored white in Fig. 2C). The explicit squeezing directions are indicated by the red arrows in Fig. 2B. Due to the transformation invariant property of the ONM [37-39], where further transformations modify only its principal axes while preserving its material parameters, the transformed shell in Fig. 2C is still composed of ONM (pink areas) but with reoriented principal axes, as indicated by the blue arrows. Detailed derivations are provided in **Supplemental Note 2**. A dimensionless parameter $M$ is defined to characterize the size of the concealed region, $M = 1 - 4\theta_M / \pi \in (0,1)$, where $\theta_M$ represents the maximum subtended angle of the concealed region relative to the coordinate center, and smaller $M$ indicates a larger concealed region. Adjusting the parameter $M$ does not affect the cloaking and concentrating functionality of the MCS. However, it alters the concealed region's geometry and reorients the principal axes of the ONM, thereby modifying the internal EM field distribution within the shell. Furthermore, varying $M$ imposes stricter fabrication precision requirements. Typically, increasing $M$ reduces the size of the concealed regions, mitigates the principal axes variation of the ONM, and eases the fabrication process of the MCS sample.

The third step involves the aperture and probe integration. A subwavelength aperture is designed to connect a concealed region with the detecting region. The probe is selected as a common microwave-frequency detector: a coaxial cable stripped of its outer cladding at one end to expose a conductive metal tip, as detailed in the inset of Fig. 2D. The aperture is designed with a diameter matching the probe's coaxial cable ($w_p$). The tip diameter ($w_t = w_p /3$) extends into the detecting region through the aperture over a distance defined as its length ($l_t$), while the opposite end of the probe remains intact as a standard coaxial cable connected to either an impedance-matched terminal (in simulations) or the sensor's main body (in experimental setups). Since the perfect wave-guiding property of the ONM relies on its structural integrity, any aperture introduced at its boundary may differently affect its perfect wave-guiding characteristics. Simulations confirm that when the aperture diameter is deep subwavelength ($w_p <= \lambda_0/20$), both the cloaking and the concentrating functions remain unaffected irrespective of the aperture-creation methods (see **Supplemental Note 3**). Therefore, under this deep-subwavelength condition, the aperture can be implemented freely without compromising performance.

### 2.2 Numerical verification of cloaking and sensitivity enhancement

After connecting the probe to the sensor's main body, we use 2D simulations to evaluate the cloaking and sensing enhancement performance of the MCS. The sensor's main body is modeled as a perfect electric conductor (PEC) structure matching the geometry of the concealed regions, while the probe is represented by a lumped port, providing a computationally efficient yet physically representative abstraction of the complete sensor system (see **Supplemental Note 4** for details). Considering the subsequent experimental fabrication, a consistent operating wavelength of $\lambda_0$ =6 cm (5 GHz) is adopted in all subsequent simulations and experiments.

The cloaking effect is verified by comparing the reduction in the scattering cross-section of sensors with MCS ($\sigma$) to those without MCS ($\sigma_0$), i.e., the ratio $\sigma/\sigma_0$. The simulated scattering incorporates both the scattering directly from the main body (the metal-bounded concealed region) and the scattering induced by subwavelength probe absorption. Figs. 2E and 2F respectively display the scattering cross-section reduction profiles as functions of probe dimensions (quantified by diameter) for $M$ =1/3 and $M$ =2/3 configurations, corresponding to the cases where the geometrical area of the sensor's main body is approximately $1.14\lambda_0^2$ and $0.57\lambda_0^2$. The simulation results in Figs. 2E and 2F indicate that when the probe size is smaller than $\lambda_0/20$, the MCS can achieve a significant 20 dB reduction in the overall sensor's scattering cross-section—regardless of whether the size of the sensor's main body is 1.14 $\lambda_0^2$ ($M$=1/3) or 0.57 $\lambda_0^2$ ($M$=2/3).

The sensitivity of a sensor is defined as the response coefficient (or amplification factor) of its output signal variation relative to input signal variation [40], which fundamentally determines its weak-signal detection capability. The proposed MCS can significantly amplify the power of the incident EM signal at the probe location, effectively multiplying the sensor's original response coefficient by an enhancement factor (EF), thereby improving its sensitivity. To quantify this improvement, we define the EF as the ratio of the EM signal power absorbed by the probe's tip with the MCS ($P$) and without the MCS ($P_0$), i.e., EF = $P/P_0$. In the simulations, the EM power absorbed by the probe's tip is transmitted via the coaxial cable to the sensor's main body located within the concealed region, with the power transfer efficiency quantifiable at the lumped port termination of the probe. The simulated EFs with varied probe dimensions are shown in Figs. 2G and 2H for $M$ =1/3 and $M$ =2/3 configurations, respectively. Since the aperture size (consistent with probe dimensions) is deeply subwavelength, different aperture-creation methods do not impact the MCS's cloaking and concentrating performance (see **Supplemental Note 3**). Thus, Figs. 2G and 2H primarily demonstrate the effect of deep subwavelength-scale probe dimensions on the EF. Sensitivity is optimized at an intermediate probe size: an overly small probe achieves weak coupling to the enhanced field, while an excessively large one introduces strong scattering, both of which degrade the overall performance. The insets in Figs. 2E-2H visually compares the field and energy density distributions with and without the MCS under conditions of satisfactory cloaking and sensing performance.

The sensor's main body in the simulations (Fig. 2) is designed on the scale of the operating wavelength ($\lambda_0$) to match experimentally fabricable dimensions. While an actual sensor body would be electrically large and impractically big to fabricate and measure at this stage, this wavelength-scale configuration suffices to validate the operating principle. Importantly, the design is scalable: the concealed regions and the sensor's main body they house can be enlarged proportionally to electrically large sizes without degrading the cloaking or sensitivity enhancement, as verified numerically in **Supplemental Note 4**.

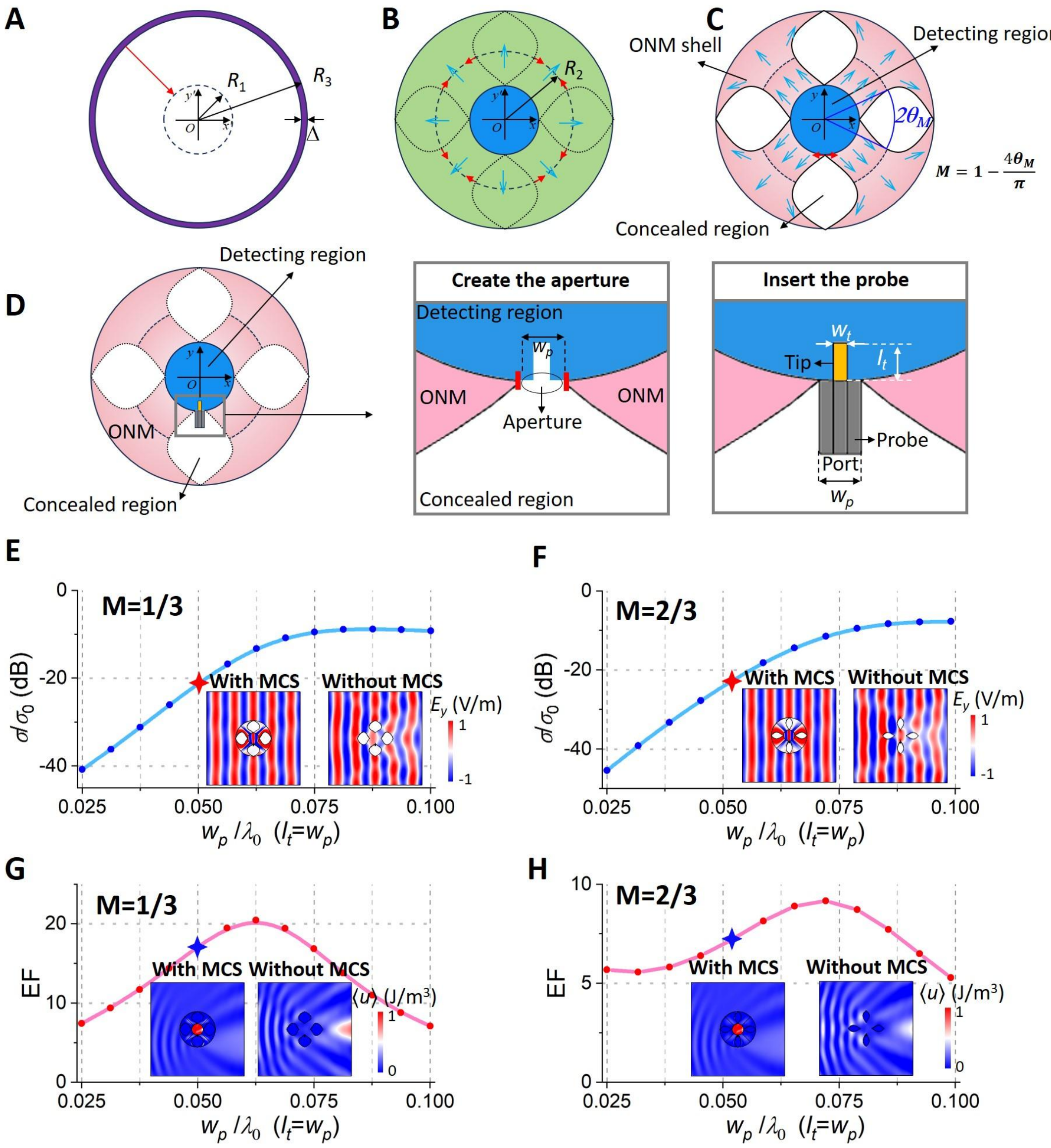

**Fig. 2. Three design steps of the MCS.** (**A**)-(**B**): design of a concentrator. (**B**)-(**C**): design of the cloak. (**C**)-(**D**): the creation of the aperture, through which the tip of the probe is inserted from the concealed region into the detecting region. (**E**)-(**F**) Scattering cross section reduction versus probe dimensions for two concealed region sizes ($M$ = 1/3, 2/3). Insets show electric field distributions (probe size marked by red stars). (**G**)-(**H**) Enhancement factor versus probe dimensions for the same $M$ values. Insets show time averaged energy density distributions (probe size marked by blue stars). All simulations are 2D cases where a detecting plane wave with the designed operating wavelength $\lambda_0$= 6 cm is incident from left to right on different structures, with ideal material parameters for the ONM (**Supplemental Note 4**) and MCS geometric dimensions of $R_1 = 0.3\lambda_0$, $R_2 = 2R_1$, and $R_3 = \sqrt{10}R_1$.

## 3 Results and discussion

### 3.1 Experimental design and sample fabrication

The transformation invariance of the ONM allows the size, number, and shape of the concealed regions to be tailored (see **Supplemental Note 4**). To simplify fabrication, we implement a design with four identical diamond-shaped regions (exhibiting 90° rotational symmetry) instead of the simulated eye-shaped ones. This configuration requires only four straight-boundary units to construct the concealed regions.

Next, considering the fabrication feasibility of subsequent sample components, the experimental validation is conducted at an operating wavelength of $\lambda_0$ =6 cm (5 GHz in the microwave band) for TM-polarized EM waves. As shown in Fig. 3A, the ONM shell is realized with subwavelength-spaced metal channel arrays filled with gradient-index media, while the HRI core is a solid high-permittivity cylinder. The sample is a 3D cylinder of height $H = 2\lambda_0$, chosen to be sufficiently tall compared to $\lambda_0$ so that its mid-plane cross-section exhibits approximately 2D behavior while remaining practical to fabricate and test. All measurements are taken at this mid-plane ($z = H/2$), where the aperture is also aligned. The geometric parameters are set to $R_1 = 0.3\lambda_0$, $R_2 = 0.6\lambda_0$, $R_3 = 0.95\lambda_0$, chosen for fabricability, measurability, and available dielectric constants ($n_c = R_3/R_1$). The fabricated MCS (Fig. 3A) comprises: (i) a 3D cylindrical ONM shell formed by 60 copper-bounded channels filled with a graded mixture of aluminum oxide ceramic and air; (ii) four diamond-shaped concealed regions embedded within the ONM shell; (iii) a solid aluminum oxide ceramic HRI core that acts as the detecting region; and (iv) an aperture linking one concealed region to the core, created by omitting the copper tape at that location. To balance cloaking (which requires $w_p \leq \lambda_0/20$) and sensitivity enhancement (optimal for $\lambda_0/40 \leq w_p \leq \lambda_0/10$), the aperture diameter is chosen as $w_p = \lambda_0/36$. A deep-subwavelength hole is drilled at the core-aperture interface to house the probe tip, with dimensions $l_t = \lambda_0/20$, and $w_t = \lambda_0/107$ (see **Supplemental Note 5** for details).

Physically, each copper-bounded subwavelength-spaced channel acts as a waveguide working at the fundamental mode [41-43], confining TM-polarized EM waves to propagate along the channel orientation. The filling fraction of aluminum oxide ceramic and air in every channel is tuned so that the effective optical path satisfies the Fabry–Pérot resonance condition, yielding zero net phase delay upon transmission. Consequently, the whole channel array operates as a practical approximation of the ONM, reproducing its wave-manipulation properties at $\lambda_0$. Moreover, the filling factor of aluminum oxide ceramic increases gradually from 0% at the outer boundary ($R_3$) to 100% at the inner boundary ($R_1$), providing impedance matching between the air background and the HRI core. This design ensures that the fabricated MCS introduces negligible scattering, thereby realizing the cloaking function.

During the fabrication of the MCS sample, both the graded aluminum oxide ceramic in the ONM shell and the solid aluminum oxide ceramic in the HRI core are manufactured using a precision computer numerical control machine. The metallic boundaries separating the subwavelength channels are realized by meticulously bonding copper tape adhesives to the edges of the aluminum oxide ceramic substrates. Additionally, a 20-mm-high base (Bambu Lab PLA Basic) with 10-mm-deep slots is produced via 3D printing to precisely accommodate the ONM shell and HRI core. After all components are fabricated, they are manually assembled into the 3D MCS sample used in the experiment (the assembly process is detailed in **Supplemental Note 6** and **Movie 2**).

**3.2 Cloaking and sensitivity enhancement performance measurement**

The cloaking performance is measured in a microwave anechoic chamber (Fig. 3B, also see **Supplemental Note 7**). In this configuration, the sensor's main body employs four PEC diamond-shaped concealed regions, consistent with the numerical study in Fig. 2. The sensor probe (a coaxial cable tip with cladding removed) is embedded within the detecting region (specifically, inside a deep-subwavelength air hole in the HRI core, perfectly conformal to the probe). The probe's opposite end passes through an aperture via coaxial cable, connects to a coaxial terminal load within a concealed region, and thereby simulates the EM wave absorption/detection effect of the sensor. The sample is illuminated by a $y$-polarized horn antenna connected to Port 1 of a vector network analyzer (VNA) and positioned 1.5 m away, providing a TM-polarized quasi-plane wave at 5 GHz. To quantify the scattering suppression, the $y$-component of the transmitted electric field ($E_y$) is scanned over a 200 mm × 200 mm area behind the sample at the mid-plane ($z = H/2$) using an automated scanning probe connected to Port 2 of the VNA (see **Supplemental Note 8** and **Movie 3** for details). Figs. 3C and 3D compare the measured (insets) and simulated (main panels) field distributions with and without the MCS. Without the MCS, strong scattering distorts the incident planar wavefront and creates a pronounced scattering shadow. In contrast, with the MCS, the transmitted wavefront remains nearly planar with no observable scattering shadow, confirming the simultaneous cloaking of both the sensor body and the actively receiving probe. The scattering reduction is quantified by the differential scattering cross-section over 4.9–5.1 GHz (Fig. 3E). The MCS achieves approximately 10 dB of reduction in simulation and 3 dB in the measurement. The measured scattering reduction (~3 dB across the band) is lower than the 2D simulated 10 dB, a discrepancy mainly arising from three factors: first, the ideal planar wave source in 2D simulations differs from the quasi-plane wave emitted by the horn antenna in experiments, which only approximates the ideal planar wave locally; second, fabrication imperfections such as dimensional deviations of 3D-printed dielectric blocks, uneven copper tape bonding (e.g., air bubbles causing irregular channel surfaces), and inherent 3D machining errors; third, material loss of aluminum oxide ceramic used in the MCS, which attenuates EM waves. Nevertheless, the clear 3 dB suppression, along with the preserved planar wavefront in the measured field patterns, provides definitive experimental proof that the MCS simultaneously suppresses scattering from both the sensor body and the active probe. Further improvement in fabrication precision and the use of lower-loss materials would narrow the gap between simulation and experiment. Furthermore, field distributions under oblique incidence (30°, 60°, and 90°) demonstrate that the cloaking effect remains robust across a wide range of incident angles (Fig. 3F).

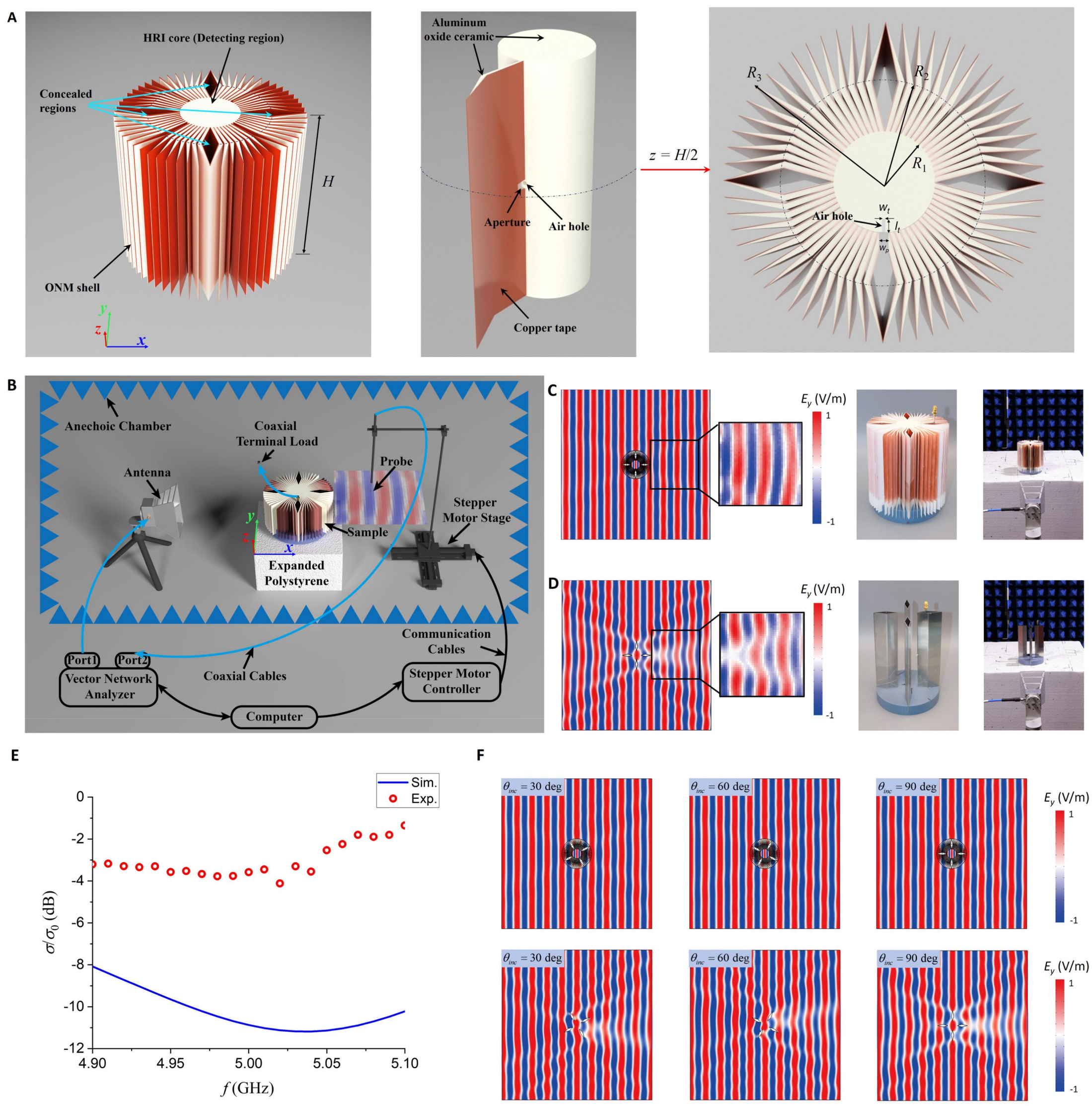

**Fig. 3. Experimental characterization of MCS cloaking performance.** (**A**) 3D schematic of the fabricated MCS sample and its structural details. (**B**) Schematic of the measurement setup. (**C**) With MCS: left panel shows simulated $E_y$ field; inset shows corresponding measured field within the outlined region; middle and right panels show the prototype and measurement photo. (**D**) Without MCS: layout same as (**C**). (**E**) Simulated and measured scattering cross section reduction near 5 GHz. (**F**) Simulated $E_y$ distributions under 30°, 60°, and 90° incidence, with MCS (top row) and without MCS (bottom row).

The sensitivity enhancement performance is quantified using the same anechoic chamber and source configuration as in the cloaking measurement, but with the scanning probe removed. Instead, the sensor probe is directly connected to Port 2 of the VNA to record the transmitted signal ($S_{21}$) (Fig. 4A; see **Supplemental Note 8 and Movie 3**). The enhancement factor is quantified using the measured transmission parameter, consistent with the definition EF = $P/P_0$ in simulations. Here, $P$ and $P_0$ are derived as $P = |S_{21(\text{with MCS})}|^2/\varepsilon_{\text{core}}$ and $P_0 = |S_{21(\text{without MCS})}|^2/\varepsilon_{\text{air}}$, where $\varepsilon_{\text{core}}$ and $\varepsilon_{\text{air}}$ denote the permittivity of aluminum oxide ceramic and air, respectively. This division by the permittivity corrects for the different dielectric environments of the probe (aluminum oxide ceramic for the MCS core vs. air for the bare sensor), as the measured $|S_{21}|^2$ is proportional to the EM energy flux density which is inherently dependent on the medium's permittivity, thus ensuring the derived $P$ and $P_0$ accurately reflect the actual EM power absorbed by the probe tip in both cases. Figs. 4B-4D show the measured power (blue and black curves) and the EF (red curves) for incident angles of 0°, 30°, and 60°. The MCS provides an approximately 8-fold power enhancement at 5.0 GHz and an average 6-fold enhancement across 4.9–5.1 GHz, with consistent performance across angles. To demonstrate that the fabricated MCS still exhibits excellent EM power enhancement performance under practical conditions, an additional demonstration experiment is conducted in a laboratory setting, where a weak background EM field is generated by a VNA. The probe of a microwave sensor equipped with a buzzer alarm is then placed in three different configurations: in free space, with the bare sensor (Fig. 3D), and with the sensor integrated with MCS (Fig. 3C and **Movie 4**). The experimental results show that the microwave sensor with the buzzer alarm does not respond to either free space or the bare sensor, but it triggers the buzzer alarm for the sensor with MCS. This further verifies that the MCS maintains excellent EM power enhancement effectiveness in real-world environments, even under random incident directions of the background field.

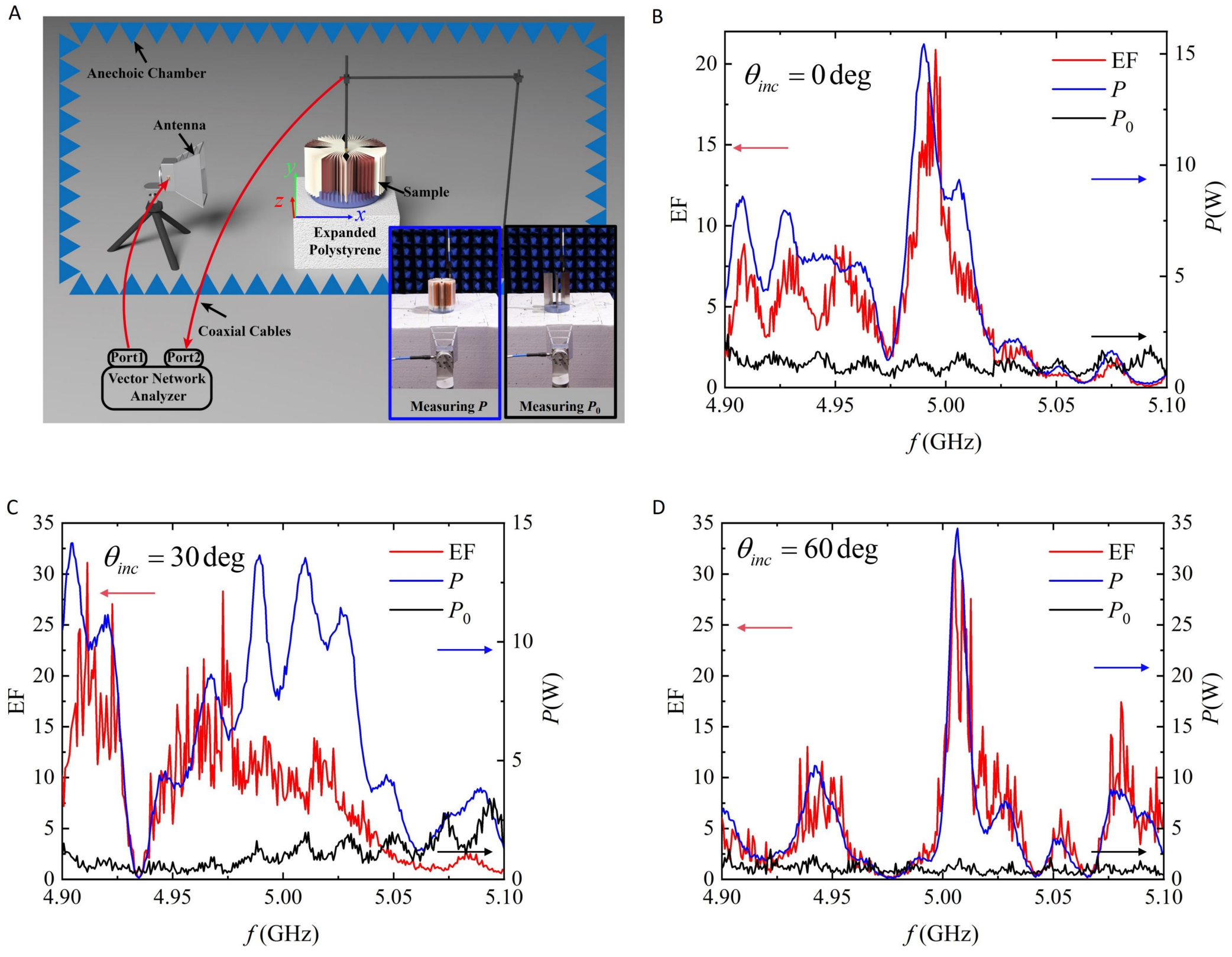

**Fig. 4. Experimental characterization of MCS sensitivity enhancement performance.** (**A**) Schematic of the measurement setup; insets show photographs of the configurations with (left) and without (right) the MCS. (**B-D**) Experimentally measured enhancement factor (EF = $P/P_0$) under (**B**) normal incidence, (**C**) 30°, and (**D**) 60° incidence. $P$ and $P_0$ are the derived powers for the MCS-coated and bare sensor, respectively, derived from the measured $|S_{21}|$.

## 4 Conclusions

In summary, we have resolved the long-standing trade-off between EM invisibility and sensing sensitivity through a system-level TO co-design strategy. Rather than treating the sensing probe and the sensor body as separate design targets, the proposed MCS architecture integrates an ONM shell, a field-concentrating HRI core, and deep-subwavelength electrical interconnections into a unified physical framework. This integrated architecture simultaneously suppresses scattering from the complete sensor system while enhancing the local EM field at the sensing probe, thereby reconciling two seemingly incompatible requirements within a single device.

Using practical ONM metamaterials, we experimentally demonstrate an invisible EM sensing platform operating at microwave frequencies ($\lambda_0$= 6 cm) that achieves broadband scattering suppression (>3 dB) together with an approximately sixfold enhancement of the detected signal. More importantly, the proposed architecture accommodates diverse sensor-body geometries, sizes, material compositions, and electrical layouts without compromising either cloaking or sensing performance, providing a practical route toward integrated invisible sensing systems.

Beyond the specific microwave implementation demonstrated here, our work establishes a general design paradigm for system-level co-optimization of wave transparency and sensing performance. The proposed framework is readily extendable to other wave-based sensing platforms governed by analogous physical principles and opens new opportunities for non-invasive weak-signal detection in biomedical diagnostics, secure communications, quantum technologies, deep-space exploration, and other applications where minimizing measurement back-action while maximizing signal reception is fundamentally desired. More broadly, this work suggests that invisibility and sensing need not be competing objectives but can instead be co-optimized through system-level wave engineering, offering a general route toward next-generation non-invasive sensing technologies.

**Acknowledgements**

We are grateful for financial supports from the National Natural Science Foundation of China (Nos. 12274317 and 12374277), San Jin Talent Support Program—Shanxi Provincial Youth Top-notch Talent Project (SJYC2025302), Natural Science Foundation of Shanxi Province (202303021211054), and Shanxi Province Higher Education Institutions Young Faculty Research and Innovation Support Program (2025Q006).